\documentclass[preprint,aps,pra,showpacs]{revtex4}
\usepackage{graphicx}
\usepackage{amsmath}
\usepackage{amsfonts}
\usepackage{bm}
\usepackage{color}

\begin{document}

\title{Coulomb glory in low-energy antiproton scattering by heavy
nucleus: screening effect of vacuum polarization}

\author{A~V~Maiorova$^1$, D~A~Telnov\footnote{telnov@pcqnt1.phys.spbu.ru}$^1$, V~M~Shabaev$^1$, G~Plunien$^2$ and
T~St\"ohlker$^{3,4}$}
\address{$^1$ Department of Physics, St.~Petersburg State
University, Ulianovskaya 1, Petrodvorets, St.~Petersburg 198504,
Russia}
\address{$^2$ Institut f\"{u}r Theoretische Physik, TU Dresden,
Mommsenstrasse 13, D-01062 Dresden, Germany}
\address{$^3$ Gesellschaft f\"{u}r Schwerionenforschung,
Planckstrasse 1, D-64291 Darmstadt, Germany}
\address{$^4$ Physikalisches Institut, Philosophenweg 12,
 D-69120 Heidelberg, Germany}

\begin{abstract}
Backward scattering of antiprotons by bare uranium is studied
theoretically for antiproton energies within the interval 100 eV
-- 1 keV. A marked maximum of the differential cross section in the
backward direction (Coulomb glory) at some energies of the incident
particle is revealed. The effect is due to the screening properties of the
vacuum polarization potential and can be regarded as a manifestation of the
vacuum polarization in non-relativistic collisions of heavy particles.
Experimental observation can become feasible with new facilities for antiproton
and ion research at GSI.
\end{abstract}
\maketitle

The project FAIR (Facility for Antiproton and Ion Research) at GSI
in Darmstadt will give an opportunity to get high-intensity
antiproton beams at energies between 30 MeV and 300 keV at a
magnetic storage ring and at energies between 300 keV and 20 keV
at an electrostatic storage ring. It will be possible to
decelerate the antiprotons to ultra-low eV energies by means of
heavy ion trap facilities. This will make accessible a large
variety of new atomic collision experiments, such as
investigations of the antiproton scattering. These investigations
open new opportunities in observation of the Coulomb glory
effect, which was predicted in Refs. \cite{gl01,gl02}. The
phenomenon results in a prominent maximum of the differential
cross section (DCS) in the backward direction at
some energy of the incident particle, provided the interaction
with the target is represented by a screened Coulomb attraction potential
(the pure Rutherford cross section has a smooth minimum at 180$^{\circ}$
irrespectively of the energy).

In the previous paper \cite{Maiorova07} we investigated the backward scattering
of antiprotons by highly charged and neutral uranium ($Z=92$). It
was shown that the Coulomb glory effect takes place due to screening of the
nuclear Coulomb attraction by the electrons. In collisions of
antiprotons with bare uranium, the Coulomb glory is also present because of the
effect of vacuum polarization (VP), which was accounted for in Ref. \cite{Maiorova07}
within the Uehling approximation.
Observation of the
Coulomb glory in collisions of antiprotons with bare uranium nuclei can
be of particular interest since the screening property of the
VP potential in non-relativistic collisions of
heavy particles can be manifested. In the present paper we study
the influence of the exact one-loop VP potential
 on the backward scattering of antiprotons with bare
uranium. The calculations have been performed in the framework of
semiclassical and quantum theory. Atomic units ($ \hbar=e=m_e=1$)
are used in the paper.

The non-relativistic scattering theory can be applied if the
kinetic energy of the antiproton is as low as a few hundreds of
electron volts. We use the partial wave expansion of the
scattering amplitude $A(\theta)$,
\begin{eqnarray}
A(\theta)=
\exp(2i\delta_{0}^{c})\left[-\frac{\nu}{2k\sin^{2}\theta/2}
\exp\left(-2i\nu \ln \sin\frac{\theta}{2}\right)\right.\nonumber \\
+ \frac{1}{k}\sum_{l=0}^{\infty} (-1)^{l}
(2l+1)P_{l}(\cos\theta)\nonumber\\
\left.\times\exp(i\delta_{l}^{s})\sin\delta_{l}^{s} \frac{(1-i
l/\nu)\dots (1-i/\nu)}{(1+i l/\nu)\dots (1+i/\nu)}\right]
\label{eq1}
\end{eqnarray}
where $k$ is the momentum of the antiproton, $\nu=-Zm_{\bar{p}}/k$
is the Coulomb parameter ($m_{\bar{p}}$ is the antiproton mass)
and $P_{l}(\cos\theta)$ are the Legendre polynomials. Note that
the total scattering amplitude is a sum of the pure Coulomb
amplitude and the contribution due to short-range (non-Coulomb)
terms in the scattering potential. The phase shifts
$\delta_{l}^{s}$ are the differences between the total phase
shifts $\delta_{l}$ and the Coulomb phase shifts $\delta_{l}^{c}$
\begin{equation}
\delta_{l}^{s}=\delta_{l}-\delta_{l}^{c}.
\end{equation}
DCS is related to the scattering amplitude as follows:
\begin{equation}
\frac{d\sigma}{d\Omega}= |A(\theta)|^{2}.
\end{equation}
The phase shifts $\delta_{l}^{s}$ can be calculated by means of
the variable phase method \cite{morse33,druk49,Babikov1} without
solving the radial Schr\"odinger equation. Within this approach,
the variable phase $\delta_{l}^{s}(k,r)$ is a solution of a first-order
differential equation. In our case, this differential equation can
be written as
\begin{eqnarray}
\frac{d}{dr}\delta_{l}^{s}(k,r)=-2m_{\bar{p}}kv(r)r^{2}\nonumber
\\
\times\left[\cos\delta_{l}^{s}(k,r)F_{l}(k,r)
-\sin\delta_{l}^{s}(k,r)G_{l}(k,r)\right]^{2}. \label{ph}
\end{eqnarray}
Here $F_{l}(k,r)$ and $G_{l}(k,r)$ are the regular and irregular
Coulomb wave functions, respectively \cite{Abramowitz}, and $v(r)$
is the short-range part of the scattering potential $V(r)$:
\begin{equation}
 v(r) = V(r) + \frac{Z}{r}.
\end{equation}
The initial condition for solving (\ref{ph}) is
$\delta_{l}^{s}(k,0)=0$. The phase shift is the limit of
$\delta_{l}^{s}(k,r)$ as $r\rightarrow\infty$:
\begin{equation}
\delta_{l}^{s} = \lim_{r\rightarrow\infty}\delta_{l}^{s}(k,r)\,.
\end{equation}
For the energies of the order of a few atomic units the Coulomb parameter is
large ($|\nu|\gg 1$), and the motion of the antiproton can be described
in the framework of the quasiclassical approximation
\cite{Landau3}. Then the phase shifts $\delta_{l}^{s}$ can be presented
as a difference of the two integrals which correspond to the
phases of the quasiclassical wave functions in the total
scattering potential and in the Coulomb potential:
\begin{eqnarray}
 \delta_{l}^{s}=\lim_{R\rightarrow\infty}\left\{\int_{R_{0}}^{R} dr
\sqrt{2m_{\bar{p}}\left(E-V(r)\right)-\frac{(l+1/2)^{2}}{r^{2}}}\right.\nonumber\\
\left. -\int_{R_{\mathrm{C}}}^{R} dr
\sqrt{2m_{\bar{p}}\left(E+\frac{Z}{r}\right)
-\frac{(l+1/2)^{2}}{r^{2}}}\right\}, \label{ph2}
\end{eqnarray}
where $R_{0}$ and $R_{\mathrm{C}}$ are the classical turning
points for the two motions, respectively. In our calculations we
applied both the variable phase method and the quasiclassical
approximation and found that the results are very close to each
other for the energy range under consideration (100 eV -- 1 keV).

The total potential $V(r)$ experienced by the antiproton due to electromagnetic
interaction is represented by a sum of the potential of a finite nucleus
and the VP potential:
\begin{equation}
V(r) = V_{\mathrm{n}} (r) + V_{\mathrm{VP}} (r)\,.
\label{th10}
\end{equation}
The potential of a finite nucleus is given by
\begin{equation}
V_{\mathrm{n}} (r) = -\int d^{3}r' \frac{\rho_{\mathrm{n}} (r')}
{|\bm{r}-\bm{r}'|}. \label{th20}
\end{equation}
Here $\rho_{\mathrm{n}}$ is the nuclear charge density, normalized to
the nuclear charge number $Z$. We employ the Fermi-like nuclear
charge distribution
\begin{equation}
\rho_{\mathrm{n}} (r) = \frac{N_{0}}{1+\exp[(r-r_{0})/a]}
\label{th30}
\end{equation}
where the parameter $a$ is equal to  $2.3/(4 \ln3)$ fm, the parameters $r_{0}$
and  $N_0$ are derived from  the root-mean-square nuclear charge radius and
the normalization condition \cite{Shabaev02}.

\begin{figure}
\includegraphics[width=\columnwidth]{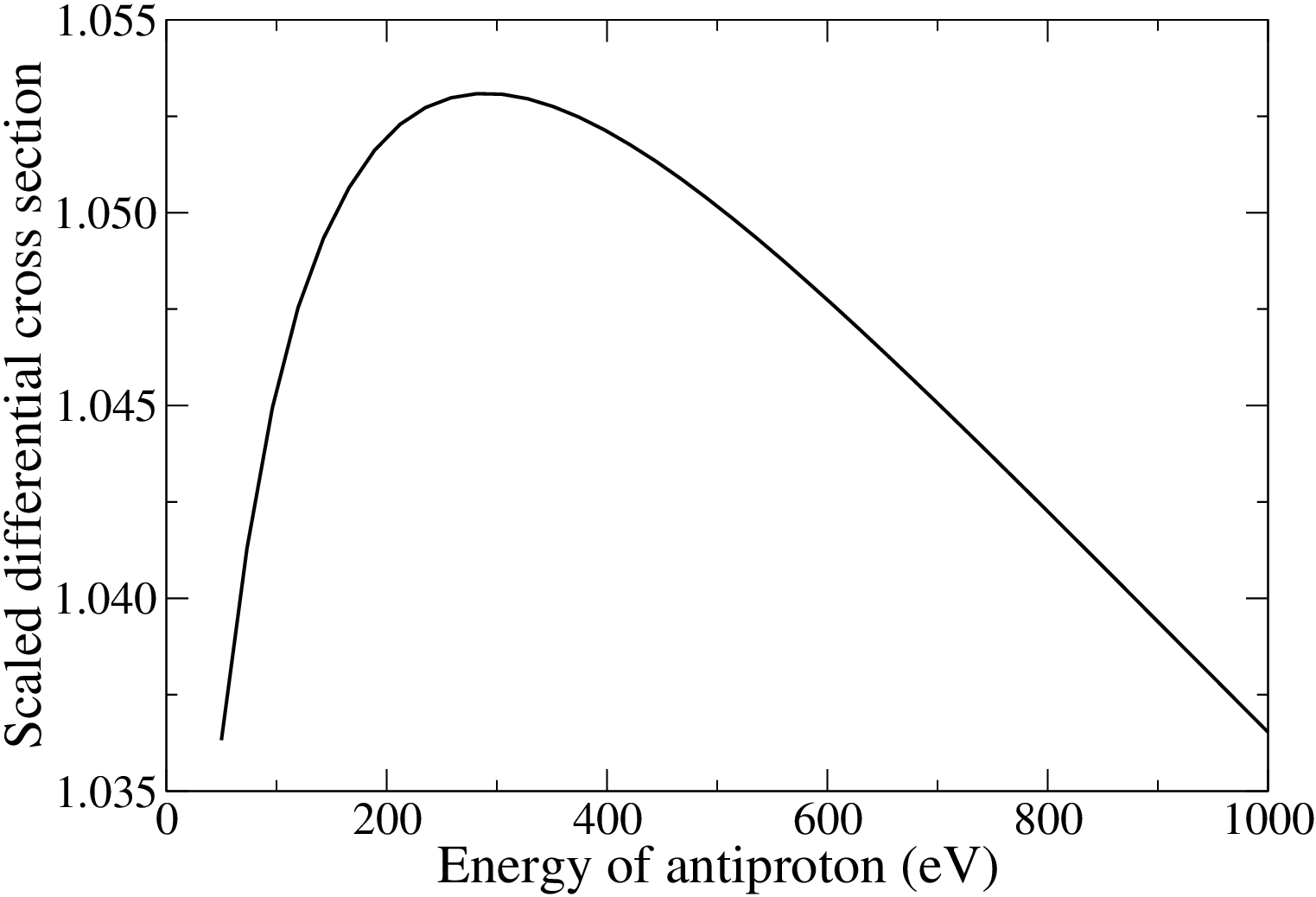}
\caption{Scaled DCS $d\tilde{\sigma}/d\Omega$ (\ref{scs}) at
180$^{\circ}$ for the total scattering
potential (\ref{th10}) vs the antiproton energy.}
\label{fig1}
\end{figure}

The VP potential is conveniently represented as a sum of the
Uehling and the Wichmann--Kroll (WK) potential:
\begin{equation}
V_{\mathrm{VP}}(r) = V_{\mathrm{Uehl}} (r) + V_{\mathrm{WK}} (r)\,.
\label{th10vp}
\end{equation}
The Uehling potential is given by the lowest-order term in the expansion
of the one-electron-loop vacuum polarization  in powers of the Coulomb
electron-nucleus interaction. According to the Furry theorem, this term contains
one Coulomb interaction in the vacuum loop and is ultraviolet divergent.
It becomes finite after charge renormalization. The renormalized expression for
the Uehling potential is given by (see, e.g., \cite{Shabaev02,Mohr98})
\begin{eqnarray}
 V_{\mathrm{Uehl}} (r)= -\frac{2}{3rc^{2}}\int_{0}^{\infty}dr'
r'\rho_{n}(r') \int_{1}^{\infty}dt\left(1+\frac{1}{2t^{2}}\right)
\frac{\sqrt{t^{2}-1}}{t^{3}}\nonumber\\
\times\left[\exp\left(-{2}{c}|r-r'|t\right)-
\exp\left(-{2}{c}(r+r')t\right)\right]
\label{th70}
\end{eqnarray}
where $c$ is the speed of light in vacuum.

The Wichmann--Kroll potential $V_{\mathrm{WK}} (r)$
accounts for the higher-order terms in the
expansion of the vacuum loop in powers of the Coulomb
electron-nucleus interaction \cite{Wichman56}.
Although the WK potential is finite, the regularization is still required
in the lowest-order non-zero term due to a spurious gauge-dependent
piece of the light-by-light scattering contribution. The spurious term
disappears if the WK potential is calculated by summing up the partial-wave
differences between the full one-loop contribution and the unrenormalized
Uehling term \cite{gyulassy75,rinker75,soff88,manakov89}.
The calculation formula for the WK potential can be written as
\cite{Mohr98,Artemmyev97}:
\begin{eqnarray}
 V_{\mathrm{WK}}(r)= \frac{2}{\pi}\sum_{\kappa=\pm
1}^{\pm\infty}|\kappa|\int_{0}^{\infty}d\omega\int_{0}^{\infty}dr_1\,
r_1^{2} \int_{0}^{\infty}dr_2\,r_2^{2}\; \frac{1}{\max(r,r_1)}\nonumber\\
\times
V_{\rm n}(r_2)\sum_{i,k=1}^{2}\mathrm{Re}\{F_{\kappa}^{ik}(i\omega,r_1,r_2)
[G_{\kappa}^{ik}(i\omega,r_1,r_2)-F_{\kappa}^{ik}(i\omega,r_1,r_2)]\}.
\label{WKpot}
\end{eqnarray}
Here $\kappa$ is the relativistic angular momentum quantum number,
$G_{\kappa}^{ik}(i\omega,y,z)$ and $F_{\kappa}^{ik}(i\omega,y,z)$ are the
radial Dirac components of the partial-wave contributions to the bound and
free electron Green's functions, respectively.

\begin{figure}
\includegraphics[width=\columnwidth]{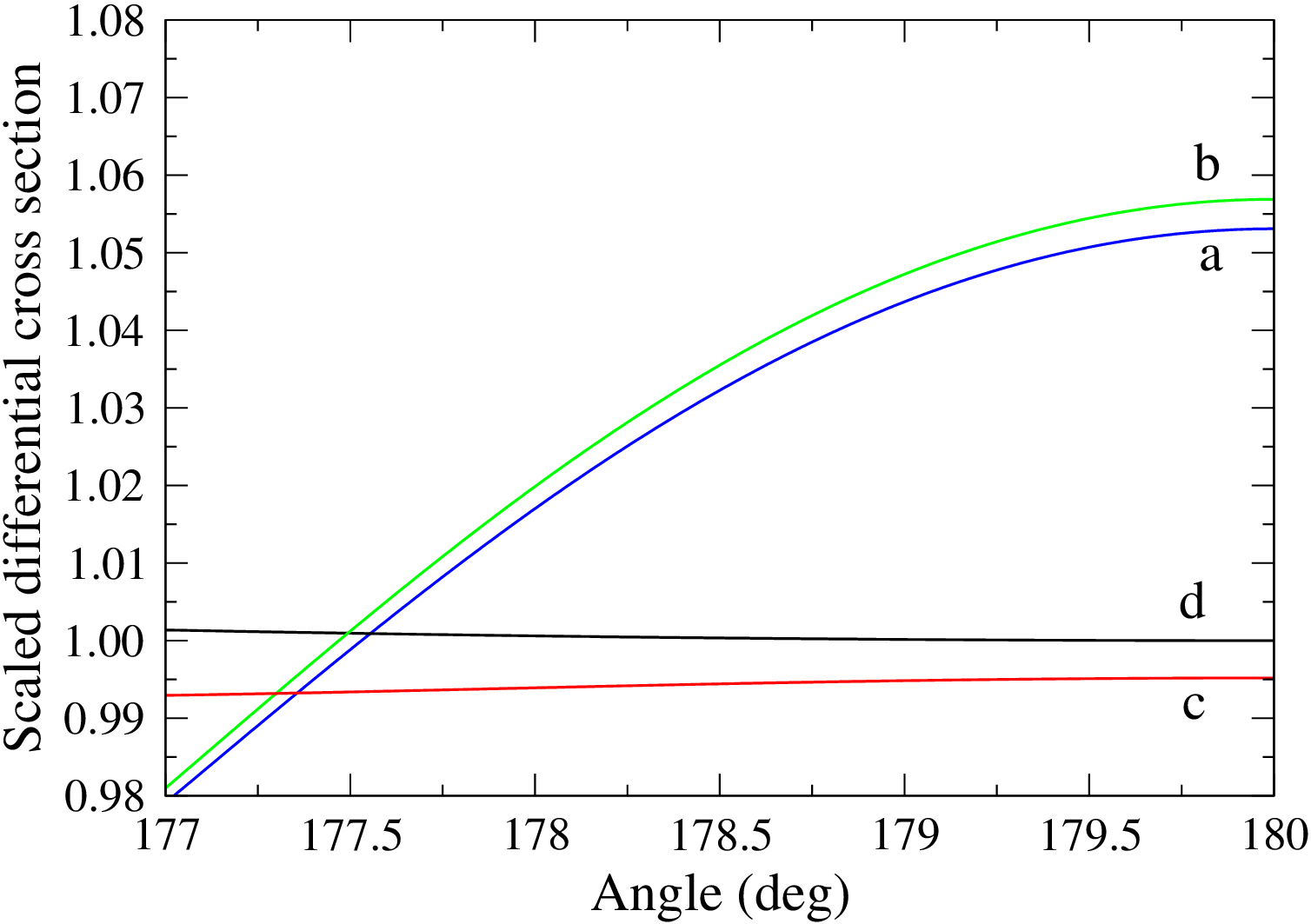}
\caption{Scaled DCS $d\tilde{\sigma}/d\Omega$ (\ref{scs})
for the energy of the antiproton 300 eV. (a) DCS for the total scattering
potential (\ref{th10}); (b) DCS for the potential including the finite nucleus
and Uehling contributions only; (c) DCS for the finite nucleus potential only;
(d) DCS for the pure Coulomb potential (scaled Rutherford cross section).}
\label{fig2}
\end{figure}
We have calculated DCS of elastic scattering of antiprotons by bare uranium
nuclei in the energy range from 100 eV to 1 keV. To make the results at
different energies comparable, the differential cross section has
been scaled according to
\begin{equation}
 \frac{d\tilde{\sigma}}{d\Omega} =
 \left(\frac{4E}{Z}\right)^{2}\frac{d\sigma}{d\Omega} .
\label{scs}
\end{equation}
The scaled Rutherford cross section does not depend on the antiproton energy and
the nuclear charge number and is equal to unity at $\theta=180^{\circ}$.
The value of $d\tilde{\sigma}/d\Omega$ at
$\theta=180^{\circ}$ represents the ratio of the DCS for the scattering by the
uranium nucleus and the corresponding Rutherford DCS and can serve as a
quantitative measure of the Coulomb glory effect.

\begin{figure}
\includegraphics[width=\columnwidth]{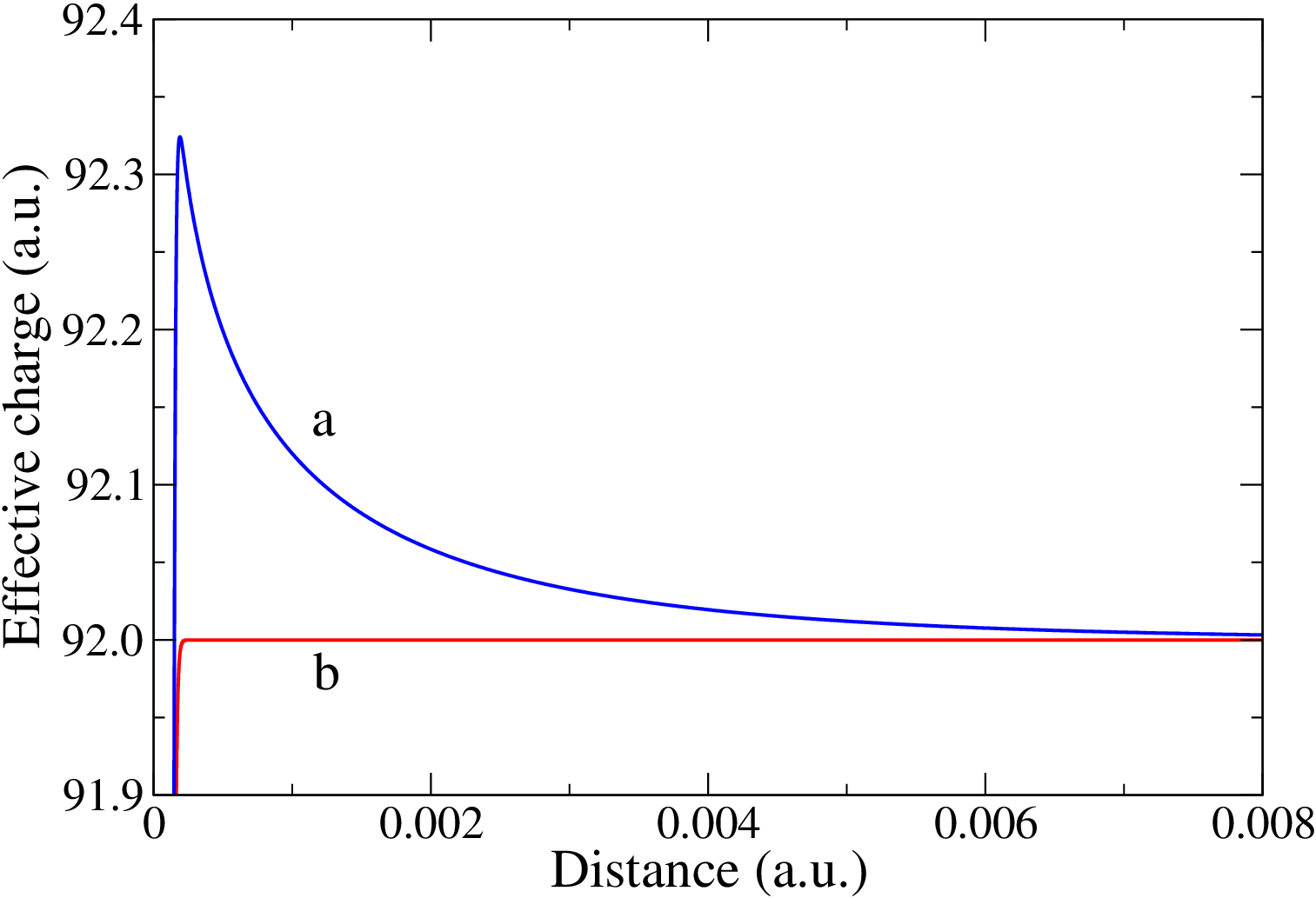}
\caption{ The effective charge  $Z_{\mathrm{eff}}(r)\equiv - rV(r)$
as a function of $r$. (a) $Z_{\mathrm{eff}}(r)$  for the total scattering
potential (\ref{th10}); (b) $Z_{\mathrm{eff}}(r)$ for the finite nucleus
potential only.}
\label{fig3}
\end{figure}
In figure~\ref{fig1} we show the scaled DCS as defined by Eq. (\ref{scs}) at
$\theta=180^{\circ}$ as a function of the antiproton energy. As one can see,
the uranium nucleus DCS is larger than the corresponding Rutherford DCS for all
the energies in the range, but the maximum Coulomb glory effect
($d\tilde{\sigma}/d\Omega (180^{\circ})=1.053$) is reached at the antiproton
energy of about 300~eV.
It should be noted that the maximum of DCS at $\theta=180^{\circ}$ is a
result of constructive interference of different angular momentum contributions
to the total scattering amplitude. The more angular momenta in equation
(\ref{eq1}) interfere constructively, the larger is the amplitude and the
cross section. At the same time, the individual contributions (that are the
phase shifts $\delta_{l}^{s}$) must not be too small. On the other hand, the
number of partial waves with large enough phase shifts depends on the range of
the scattering potential. That is why the three terms in the total scattering
potential (\ref{th10}) are not equally important for the Coulomb glory effect.
To estimate the influence of the different terms in the total
scattering potential (\ref{th10}) on the Coulomb glory, we have also calculated
DCS with partial scattering potentials. For the antiproton energy 300 eV, the
DCS dependences on the scattering angle are presented in figure~\ref{fig2}.
A significant deviation of the finite nucleus potential from the pure Coulomb
potential exists at very small distances which have the order of the nuclear
size. In the scattering amplitude it affects about seven first  angular momenta
only. Thus the DCS for the finite nucleus potential is close to the Rutherford
cross section and does not show any maximum at $180^{\circ}$. For the same
reason, the strong interaction between the antiproton and the nucleus is not
important for the Coulomb glory effect either, and we include only
electromagnetic interaction in the scattering potential (\ref{th10}).
As one can see from figure~\ref{fig2}, the most important part of the total
scattering potential is
the Uehling potential which is responsible for the DCS maximum at $180^{\circ}$.
Although the Wichmann--Kroll potential has a longer range than the Uehling
potential, its absolute value is much smaller. The effect of the Wichmann--Kroll
potential results in a slight decrease of the DCS in the vicinity of
$\theta=180^{\circ}$. We emphasize that in the case under consideration
the Coloumb glory effect results exclusively from the screening property of the
VP potential. This property means a decrease of the effective charge
$Z_{\mathrm{eff}}(r)\equiv -rV(r)$ with $r$ increasing. In figure~\ref{fig3}
we plot the effective charge $Z_{\mathrm{eff}}$ versus the distance $r$ for the
total scattering potential as well as for the finite nucleus potential only.
Except for the small region inside the nucleus, the effective charge for the
total potential smoothly decreases from the maximum value of $92.32$ to the
uranium charge number $92$. For the finite nucleus potential, the effective
charge quickly approaches the value $92$ just outside the nucleus.

In summary, we have investigated the elastic scattering of low-energy
antiprotons by bare uranium nuclei at large angles and found that
the differential cross section has a maximum in the backward direction (the
Coulomb glory). The largest effect is predicted for the antiproton energy
of about 300 eV. The existence  of the Coulomb glory requires a screened
Coulomb potential and, therefore, the phenomenon revealed is due to the screening
properties of the vacuum polarization potential. Possible experimental
observation of the Coulomb glory in collisions of antiprotons with
bare uranium nuclei at the new GSI facility can become a clear manifestation of
the vacuum polarization effects in non-relativistic collisions of
heavy particles.

\acknowledgments
 This work was partially supported by INTAS-GSI
(Grant No. 06-1000012-8881), RFBR (Grant No. 07-02-00126a), DFG
and GSI. A.V.M. also acknowledges the support from  DAAD and the
Dynasty foundation.


\end{document}